\documentclass[pra,twocolumn,showpacs,preprintnumbers,amsmath,amssymb]{revtex4}

\usepackage{natbib}
\usepackage{graphicx}% Include figure files
\usepackage{dcolumn}% Align table columns on decimal point
\usepackage{bm}% bold math
\newcommand{\bra}[1]{\langle#1|}

\newcommand{\scal}[2]{\langle#1|#2\rangle}\newcommand{\ket}[1]{|#1\rangle}
\begin{document}

\preprint{APS/123-QED}

\title{Generating and Revealing a Quantum Superposition of Electromagnetic Field Binomial States in a Cavity}

\author{R. Lo Franco}
 \email{lofranco@fisica.unipa.it}
 \homepage{http://www.fisica.unipa.it/~lofranco}
\author{G. Compagno}
  \author{A. Messina}
 \author{A. Napoli}

\affiliation{Dipartimento di Scienze Fisiche ed Astronomiche,
Universit\`{a} di Palermo, via Archirafi 36, 90123 Palermo, Italy}

\date{\today}% It is always \today, today,

\begin{abstract}
We introduce the $N$-photon quantum superposition of two orthogonal
generalized binomial states of electromagnetic field. We then
propose, using resonant atom-cavity interactions, non-conditional
schemes to generate and reveal such a quantum superposition for the
two-photon case in a single-mode high-$Q$ cavity. We finally discuss
the implementation of the proposed schemes.
\end{abstract}


\maketitle

Since the birth of the Schr\"odinger cat phenomenon \cite{schro},
the possibility of generating and detecting macroscopic quantum
superpositions has been holding much interest in several frameworks
\cite{wheel,jeong1,caval}. A macroscopic quantum superposition of
electromagnetic field states is usually meant as a superposition of
two coherent states with classically different phases
\cite{har,haroche}. In the context of cavity quantum electrodynamics
(CQED), such a state has been generated by dispersive coupling
between a circular Rydberg atom and a small coherent cavity field,
the quantum decoherence of the superposition being there observed by
probe atoms \cite{brune1}. Other schemes have been proposed to
generate or detect quantum superpositions of this kind, for example
in a dispersive medium \cite{yurke}, in a nanomechanical resonator
\cite{tian} and in a cavity \cite{baseia,liu,vid2,casagrande} and a
free-propagating light pulse was also recently prepared in such a
state \cite{ourj}. Nevertheless, two different coherent states can
never be made exactly orthogonal and therefore the coherent states
of a quantum superposition are not completely distinguishable. In
the CQED experiment of Ref.~\cite{brune1}, it is for example
necessary to adjust the detuning between the atomic transition and
the cavity frequency to partially distinguish the two components of
the superposition. Thus, to propose schemes aimed at generating
``optimized'' quantum superpositions, defined as quantum
superpositions of two orthogonal, distinguishable electromagnetic
field states with nonzero mean fields, appears to be an attractive
challenge.

It is well known that the binomial states of electromagnetic field
are characterized by a finite maximum number of photons $N$,
interpolate between coherent state and number state and also exhibit
nonzero mean fields \cite{sto,vid,vid1}. In addition, it is always
possible to find orthogonal couples among the $N$-photon generalized
binomial states \cite{lof}. Such features make the generalized
binomial states promising candidates to construct optimized quantum
superpositions and to study the general problem of classical-quantum
border and quantum measurement. The point is then on how to generate
and reveal such states. This paper addresses this issue, by
exploiting standard resonant atom-cavity interactions in the CQED
framework.

The dynamics of the resonant interaction between a two-level atom
and a single-mode cavity field of frequency $\omega$ is described by
the Jaynes-Cummings Hamiltonian
$H_{JC}=\hbar\omega\sigma_z/2+\hbar\omega a^{\dag}a+i\hbar
g(\sigma_+a-\sigma_-a^{\dag})$, where $a$ and $a^{\dag}$ are the
field annihilation and creation operators,
$\sigma_z=\ket{\uparrow}\bra{\uparrow}-\ket{\downarrow}\bra{\downarrow}$,
$\sigma_+=(\sigma_-)^\dag=\ket{\uparrow}\bra{\downarrow}$ the
pseudo-spin atomic operators, $\ket{\uparrow}$ and
$\ket{\downarrow}$ being respectively the excited and ground state
of the two-level atom, and $g$ is the atom-field coupling constant.
The $H_{JC}$ based time evolution of the states $\ket{\uparrow
n}\equiv\ket{\uparrow}\ket{n}$ and $\ket{\downarrow
n}\equiv\ket{\downarrow}\ket{n}$, with $a^\dag a\ket{n}=n\ket{n}$,
is \cite{scully} {\setlength\arraycolsep{2pt}\begin{eqnarray}
\ket{\uparrow n}&\rightarrow&\cos(g\sqrt{n+1}t)\ket{\uparrow n}-\sin(g\sqrt{n+1}t)\ket{\downarrow n+1},\nonumber\\
\ket{\downarrow n}&\rightarrow&\cos(g\sqrt{n}t)\ket{\downarrow
n}+\sin(g\sqrt{n}t)\ket{\uparrow n-1},\label{evo}
\end{eqnarray}}
where $t$ is the atom-cavity interaction time.

\begin{figure}
\includegraphics[height=0.13\textheight, width=0.4\textwidth]{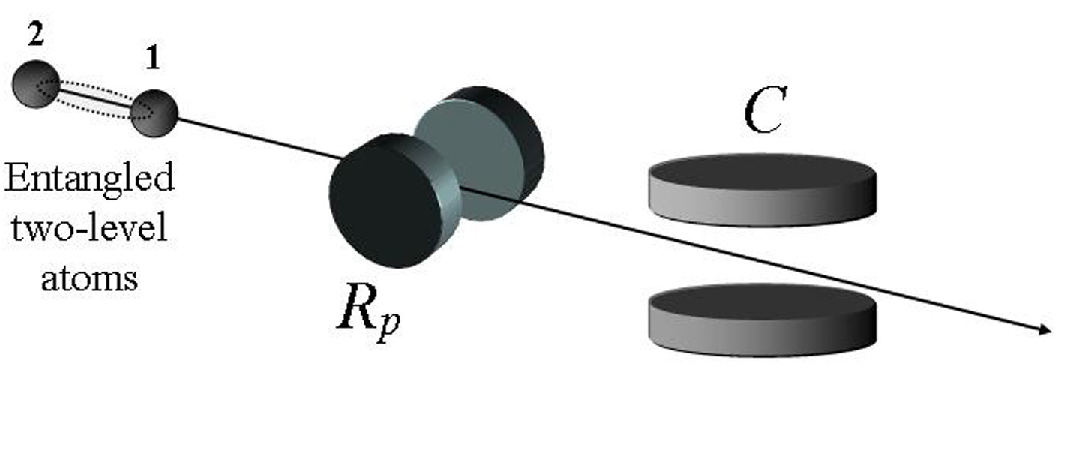}
\caption{\label{fig1}Experimental scheme for the generation of a
2QSB. $R_p$ is the ``preparing'' Ramsey zone.}
\end{figure}

\begin{figure}
\includegraphics[height=0.13\textheight, width=0.46\textwidth]{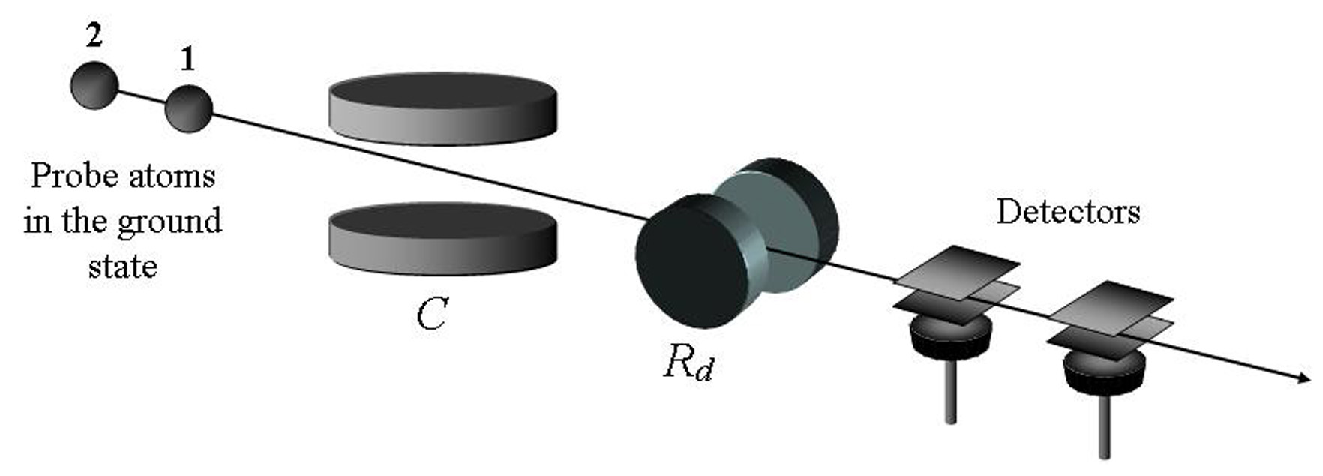}
\caption{\label{fig2}Experimental scheme for distinguishing the two
components of the 2QSB. $R_d$ is the ``decoding'' Ramsey zone.}
\end{figure}

\begin{figure}
\includegraphics[height=0.14\textheight, width=0.46\textwidth]{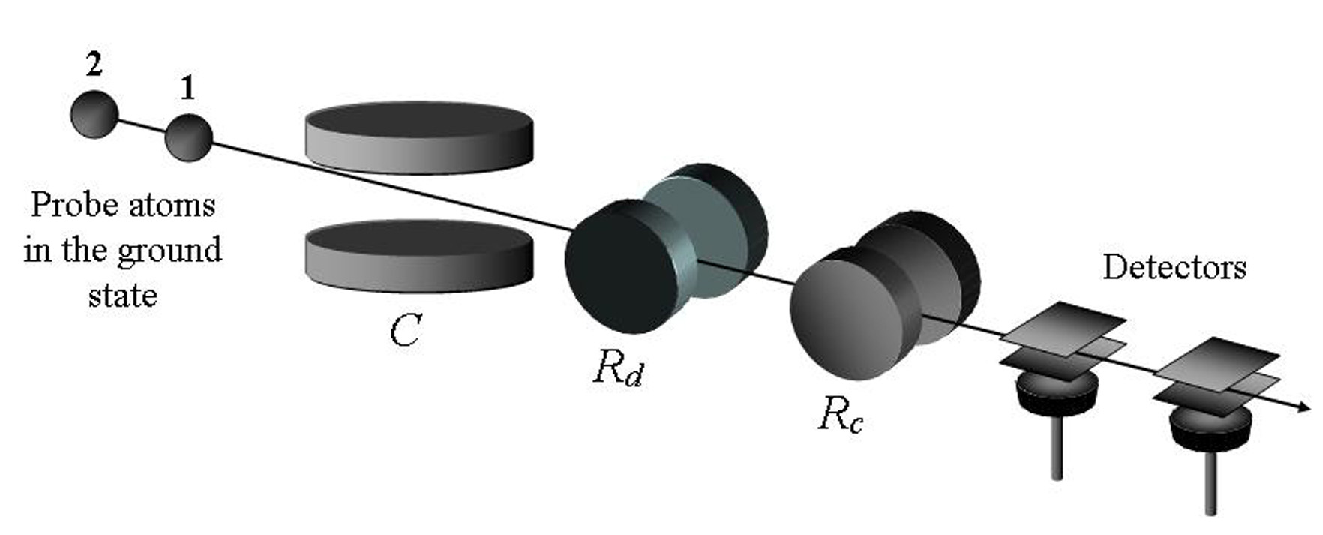}
\caption{\label{fig3}Experimental scheme for revealing the coherence
of the 2QSB. $R_c$ is the ``coherence decoding'' Ramsey zone. The
cavity $C$ is prepared in the 2QSB
$\ket{\Psi_\textrm{S}^{(2)}}_\pm$.}
\end{figure}

The normalized $N$-photon generalized binomial state is given by
\cite{sto}
\begin{equation}
\ket{N,p,\phi}=\sum_{n=0}^N\left[{N\choose
n}p^{n}(1-p)^{N-n}\right]^{1/2}e^{in\phi}\ket{n},\label{bin}
\end{equation}
where $0\leq p\leq1$ is the probability of single photon occurrence
and $\phi$ is the mean phase \cite{vid}. The orthogonality property
$\scal{N,p,\phi}{N,1-p,\pi+\phi}=0$ \cite{lof} allows us to define
the $N$-photon quantum superposition of two orthogonal generalized
binomial states ($N$QSB) as
\begin{equation}
\ket{\Psi_\textrm{S}^{(N)}}\equiv\mathcal{N}[\ket{N,p,\phi}+\eta\ket{N,1-p,\pi+\phi}],\label{bsc}
\end{equation}
where $\eta$ is a complex number and
$\mathcal{N}=1/\sqrt{1+|\eta|^2}$. It should be noted that, for
$p=0,1$, the $N$QSB is reduced to a quantum superposition of the
number states $\ket{0},\ket{N}$. The $N$QSB effectively represents a
macroscopic superposition of electromagnetic field states if
$N\gg1$. However, in order to remain in the grasp of the current
experimental feasibility, we shall concentrate on both the
generation and the revealing of the quantum superposition in the
case $N=2$. We shall show that the 2QSB
$\ket{\Psi_\textrm{S}^{(2)}}$ may be generated in a cavity by the
experimental scheme sketched in Fig.~\ref{fig1} and its components
and coherence may be revealed by the schemes sketched in
Fig.~\ref{fig2} and Fig.~\ref{fig3}.

\textit{Generating the quantum superposition}.---In the generation
scheme of Fig.~\ref{fig1}, the cavity $C$ is initially prepared in
the vacuum state $\ket{0}$, and a couple of two-level atoms, namely
1 and 2, is prepared in the state
$\ket{\psi}=\mathcal{N}(\ket{\uparrow_1\downarrow_2}+\eta_0\ket{\downarrow_1\uparrow_2})$,
with $\eta_0$ real. Entangled atomic states of this form have
already been obtained using a cavity as atomic entanglement catalyst
\cite{rai,hag}, providing in addition that the two atoms enter the
Ramsey zone, as well as the cavity, one at a time. Each atom first
crosses a ``preparing'' Ramsey zone $R_p$. The Ramsey zone
interaction makes the $j$-th atom undergo the following
transformations:
\begin{eqnarray}
\ket{\uparrow_j}&\stackrel{R}{\rightarrow}
&\cos(\theta_j/2)\ket{\uparrow_j}-e^{i\varphi_j}\sin(\theta_j/2)\ket{\downarrow_j},\nonumber\\
\ket{\downarrow_j}&\stackrel{R}{\rightarrow} &
e^{-i\varphi_j}\sin(\theta_j/2)\ket{\uparrow_j}+\cos(\theta_j/2)\ket{\downarrow_j},\label{ramseyeq}
\end{eqnarray}
where the parameters $\theta_j$ (``Ramsey pulse'') and $\varphi_j$
are fixed by adjusting the classical field amplitude and the
atom-field interaction time. The $j$-th atom then resonantly
interacts with $C$ for a time $T_j$ ($j=1,2$). The atom-cavity
interaction times $T_j$ can be obtained by selecting either
different velocities for each atom or the same velocity for the two
atoms (``monokinetic atomic beam'') and applying a Stark shift
inside the cavity for a time such as to have the desired resonant
interaction time \cite{hag,davi1}. The appropriate atomic velocity
may be selected by laser induced atomic pumping \cite{har1}. We
shall show that a 2QSB state can be efficiently generated by
appropriately choosing the Ramsey zone settings and the atom-cavity
interaction times.

In accordance with the scheme of Fig.~\ref{fig1}, atom 1 crosses
$R_p$ set with a ``pulse'' $\theta_1$ such that
$\cos(\theta_1/2)\equiv\sqrt{p}$,
$\sin(\theta_1/2)\equiv\sqrt{1-p}$, and with $\varphi_1$ to be
related to the mean phase $\phi$ appearing in Eq.~(\ref{bsc}). After
a free evolution time $\tau_1$ between $R_p$ and $C$, atom 1
interacts with the cavity $C$ for a given time $T_1$. After it exits
$C$, atom 2 crosses the Ramsey zone $R_p$, freely evolves for a time
$\tau_2$ from $R_p$ to $C$ and then interacts with $C$ for a time
$T_2$. Let us indicate with $T$ the time elapsed between the exit of
the atom 1 from $C$ and the entrance of the atom 2 in $C$. After the
passage of atom 1, the $R_p$ parameters must be reset at
$\theta_2=\theta_1+\pi$, so that $\cos(\theta_2/2)=-\sqrt{1-p}$ and
$\sin(\theta_2/2)=\sqrt{p}$ in Eq.~(\ref{ramseyeq}), and
$\varphi_2=\varphi_1+\omega(\tau_1+T-\tau_2)$. Taking into account
Eqs.~(\ref{evo}) and (\ref{ramseyeq}) it is possible to demonstrate
that, if $T_1=(4m+1)\pi/2g$ ($m$ non-negative integer) and $T_2$ is
such that the two equalities $\sin(gT_2+\pi/4)=1$,
$\sin(g\sqrt{2}T_2)=1$ are simultaneously satisfied, when the second
atom leaves $C$, the state of the total system
(atom~1+atom~2+cavity) turns out to be factorized as
$\ket{\Psi_\textrm{S}^{(2)}}\ket{\downarrow_1\downarrow_2}$. It is
worth noting that, choosing $T_2=41\pi/4g$, both equalities above
are satisfied within the error due to the typical experimental
interaction time uncertainties \cite{lof2}. Thus, the cavity field
after the passage of the two atoms coincides with the quantum
superposition of a couple of orthogonal two-photon generalized
binomial states (2QSB)
\begin{equation}
\ket{\Psi_\textrm{S}^{(2)}}=\mathcal{N}[\ket{2,p,\phi}+\eta_0
e^{i\gamma}\ket{2,1-p,\pi+\phi}],\label{2bsc}
\end{equation}
where $\phi=-[\varphi_1+\omega(\tau_1+T)]$ and
$\gamma=\omega(t_{R_2}-t_{R_1}-T_1)$, $t_{R_1},t_{R_2}$ being
respectively the interaction times of the atoms 1 and 2 with $R_p$.
It is of relevance that our procedure to generate a 2QSB in a cavity
does not require a final atomic measurement and then it is a
non-conditional scheme.

We shall now analyze the possibility to probe the generated
$\ket{\Psi_\textrm{S}^{(2)}}$ state. Generally speaking, to probe a
quantum superposition requires a measurement procedure permitting
both to resolve the two components and to reveal their relative
quantum coherence. In the following, we present a procedure
appropriate for the 2QSB $\ket{\Psi_\textrm{S}^{(2)}}$ based on
two-level probe atoms that ``read'' the cavity field. Our
considerations will be developed for the maximal 2QSB of
Eq.~(\ref{2bsc}), corresponding to $\eta_0=\pm1$, that is
\begin{equation}
\ket{\Psi_\textrm{S}^{(2)}}_\pm=[\ket{2,p,\phi}\pm
e^{i\gamma}\ket{2,1-p,\pi+\phi}]/\sqrt{2}.\label{2bscmax}
\end{equation}

\textit{Revealing the two components}.---The experimental scheme we
propose is illustrated in Fig.~\ref{fig2}. It exploits two
consecutive probe atoms both in their ground state interacting one
at a time with the apparatus. The atom 1 resonantly interacts with
$C$ for an appropriate time $T_{P_1}$, after a delay time $t_1$ from
$C$ to $R_d$ it crosses the ``decoding'' Ramsey zone $R_d$ and it is
finally measured by field ionization detectors. After this
measurement, atom 2 enters the cavity $C$. Let us denote with $T'$
the time interval between the exit of the atom 1 from $C$ and the
entrance of the atom 2 in $C$. Atom 2 resonantly interacts with $C$
for a time $T_{P_2}$, takes a time $t_2$ to go from $C$ to $R_d$,
crosses $R_d$ and its internal state is finally measured.

Let us suppose the cavity prepared in the state $\ket{2,p,\phi}$
($\ket{2,1-p,\pi+\phi}$) and perform the experiment previously
described fixing $T_{P_1}=41\pi/4g$, $T_{P_2}=(4m+1)\pi/2g$, the
$R_d$ parameters $\theta_{d_1}=\theta_{d_2}=\theta_d$ such that
$\cos(\theta_d/2)=\sqrt{p}$, $\sin(\theta_d/2)=\sqrt{1-p}$ and
$\varphi_{d_1}=-\phi+\omega t_1$,
$\varphi_{d_2}=-\phi+\omega(T'+t_2)$. Under these conditions, the
probability of finding the two atoms in the states
$\ket{\uparrow_1\uparrow_2}$ ($\ket{\downarrow_1\downarrow_2}$) at
the end of the experiment is equal to one. This statement readily
follows from the unitary evolutions \cite{lof,lof2}
\begin{eqnarray}
\ket{\downarrow_1}\ket{2,p,\phi}&\stackrel{T_{P_1},R_d}{\longrightarrow}&e^{-i\varphi_{d_1}}\ket{1,p,\phi'}\ket{\uparrow_1},\nonumber\\
\ket{\downarrow_2}\ket{1,p,\phi'}&\stackrel{T_{P_2},R_d}{\longrightarrow}&e^{-i\varphi_{d_2}}\ket{0}\ket{\uparrow_2},\label{probe1}
\end{eqnarray}
and
\begin{eqnarray}
\ket{\downarrow_1}\ket{2,1-p,\pi+\phi}&\stackrel{T_{P_1},R_d}{\longrightarrow}&\ket{1,1-p,\pi+\phi'}\ket{\downarrow_1},\nonumber\\
\ket{\downarrow_2}\ket{1,1-p,\pi+\phi'}&\stackrel{T_{P_2},R_d}{\longrightarrow}&\ket{0}\ket{\downarrow_2},\label{probe2}
\end{eqnarray}
where $\ket{1,p,\phi'}$ is the one-photon generalized binomial state
as given by Eq.~(\ref{bin}) and $\phi'=\phi-\omega T'$. Thus, we
claim that, if the only possible outcomes of our experiment are
$\ket{\uparrow_1\uparrow_2}$ or $\ket{\downarrow_1\downarrow_2}$,
then the cavity is with certainty in a combination (quantum or
statistical) of the two generalized binomial states
$\ket{2,p,\phi}$, $\ket{2,1-p,\pi+\phi}$. In a sense, the atoms act
here as ``quantum probes'', changing their state if they find the
state $\ket{2,p,\phi}$, or maintaining the same state if they find
the state $\ket{2,1-p,\pi+\phi}$.

\textit{Revealing the quantum coherence}.---Our goal of revealing
with certainty the existence of the field state
$\ket{\Psi_\textrm{S}^{(2)}}_+$ or $\ket{\Psi_\textrm{S}^{(2)}}_-$
of Eq.~(\ref{2bscmax}) inside the cavity requires a further step.
The previous experimental scheme does not indeed enable us to
discover the quantum nature of the superposition of the two
components $\ket{2,p,\phi}$, $\ket{2,1-p,\pi+\phi}$. To this end we
propose a further experimental scheme after the previous one. The
appropriate apparatus is sketched in Fig.~\ref{fig3} and differs
from the one in Fig.~\ref{fig2} for the presence of a second Ramsey
zone $R_c$, playing the peculiar role of decoding the information
about the quantum coherence we are seeking. Once again, the two
probe atoms are initially prepared in their ground state and
interact one at a time with the apparatus. In this second scheme,
the atom 1 (2) crosses the cavity $C$, the Ramsey zone $R_d$ and,
after a free evolution time $t'_1$ ($t'_2$) it enters the additional
Ramsey zone $R_c$. The procedure ends measuring the internal states
of both probe atoms after atom 2 exits $R_c$. The atom-cavity
interaction times $T_{P_1},T_{P_2}$ and the $R_d$ parameters
$\theta_d,\varphi_{d_1},\varphi_{d_2}$ are the same of the previous
scheme, and we strategically set the $R_c$ pulse $\theta_c=\pi/2$.
In order to find the suitable value of the $R_c$ parameter
$\varphi_c$, it is worth to note that, with these experimental
settings, after atom 2 leaves $R_c$ the initial total state
$\ket{\downarrow_1\downarrow_2}\ket{\Psi_\textrm{S}^{(2)}}_\pm$
evolves into
\begin{eqnarray}
\ket{0}\bigg\{\frac{1\pm
e^{i(\gamma-\phi_t-2\varphi_c)}}{2\sqrt{2}}[\ket{\uparrow_1\uparrow_2}+e^{i\alpha}\ket{\downarrow_1\downarrow_2}]\nonumber\\
-e^{i\varphi_c}\frac{1\mp
e^{i(\gamma-\phi_t-2\varphi_c)}}{2\sqrt{2}}[\ket{\uparrow_1\downarrow_2}+e^{i\beta}\ket{\downarrow_1\uparrow_2}]\bigg\},\label{probeint}
\end{eqnarray}
where $\ket{0}$ is the cavity vacuum state,
$\phi_t=2\phi-\omega(T'+t_1+t'_1+t_2+t'_2)$,
$\alpha\equiv2\varphi_c+\omega\bar{t}$ and
$\beta\equiv\omega\bar{t}$, with $\bar{t}$ being the sum of some
characteristic times of the procedure. From Eq.~(\ref{probeint}) it
is readily seen that, setting $\varphi_c=(\gamma-\phi_t)/2$, the
following unitary evolutions are obtained:
\begin{eqnarray}
\ket{\downarrow_1\downarrow_2}\ket{\Psi_\textrm{S}^{(2)}}_+&\rightarrow&
\ket{0}[\ket{\uparrow_1\uparrow_2}+e^{i\alpha}\ket{\downarrow_1\downarrow_2}]/\sqrt{2},\nonumber\\
\ket{\downarrow_1\downarrow_2}\ket{\Psi_\textrm{S}^{(2)}}_-&\rightarrow&
\ket{0}[\ket{\uparrow_1\downarrow_2}+e^{i\beta}\ket{\downarrow_1\uparrow_2}]/\sqrt{2}.\label{probe3}
\end{eqnarray}
Note that all the free evolution times can be determined from the
atomic velocities, the delay time $T_0$ between the two atoms and
the geometrical parameters. Eq.~(\ref{probe3}) says that the unitary
evolution of the probe atoms and the cavity field
$\ket{\Psi_\textrm{S}^{(2)}}_+$ ($\ket{\Psi_\textrm{S}^{(2)}}_-$)
generates a vanishing probability amplitude for the atomic states
$\ket{\uparrow_1\downarrow_2}$ and $\ket{\downarrow_1\uparrow_2}$
($\ket{\uparrow_1\uparrow_2}$ and $\ket{\downarrow_1\downarrow_2}$),
equally distributing the probability between the other two possible
outcomes $\ket{\uparrow_1\uparrow_2}$ and
$\ket{\downarrow_1\downarrow_2}$ ($\ket{\uparrow_1\downarrow_2}$ and
$\ket{\downarrow_1\uparrow_2}$). Therefore, after repeating this
experiment many times, including the preparation of the cavity
field, we are able to confirm the quantum coherence (``sign'' and
relative phase $\gamma$) of the initial cavity field state
$\ket{\Psi_\textrm{S}^{(2)}}_+$ or $\ket{\Psi_\textrm{S}^{(2)}}_-$.
In fact, if the outcomes of the repeated measurements always give
``parallel'' atoms then the cavity field is with certainty in the
quantum superposition $\ket{\Psi_\textrm{S}^{(2)}}_+$; otherwise, if
the outcomes always give ``antiparallel'' atoms then the cavity
field is with certainty in the quantum superposition
$\ket{\Psi_\textrm{S}^{(2)}}_-$.

We now briefly analyze the experimental feasibility of the proposed
schemes. They require precise atom-cavity interaction times.
However, the experimental uncertainty of the selected velocity
$\Delta v$ induces an error $\Delta T$ on the interaction time such
that $\Delta T/T\approx\Delta v/v$. In current laboratory
experiments it is possible to select a given atomic velocity such
that $\Delta v/v\leq10^{-2}$ \cite{hag,har1}. This error does not
appear to sensibly affect our schemes. We have also ignored the
atomic or photon decay during the atom-cavity interactions. This
assumption is valid if $\tau_\textrm{at},\tau_\textrm{cav}>T$, where
$\tau_\textrm{at},\tau_\textrm{cav}$ are the atomic and photon mean
lifetimes respectively and $T$ is the interaction time. For circular
Rydberg atomic levels and microwave superconducting cavities with
quality factors $Q\sim10^8-10^{10}$ the required inequality on the
mean lifetimes can indeed be satisfied, being
$\tau_\textrm{at}\sim10^{-5}-10^{-2}\textrm{s}$,
$\tau_\textrm{cav}\sim10^{-4}-10^{-1}\textrm{s}$ and
$T\sim10^{-5}-10^{-4}\textrm{s}$ \cite{har,rai}. Moreover, the
typical mean lifetimes of circular Rydberg atomic levels
$\tau_\textrm{at}$ are such that the atoms do not decay during the
entire sequence of the schemes \cite{rai,har1}. The delay time $T_0$
between the two atoms can be adjusted so that they cross the
experimental apparatus one at a time, as required by our schemes.
Recent laboratory developments open promising perspectives for a
better and easy control of a well-defined atom numbers sequence
\cite{maioli} and for a high efficiency atomic detection in
microwave CQED experiments \cite{auff}.

Finally, because the binomial states interpolate between number and
coherent states, an estimate of the time scale of 2QSB decoherence
can be provided by the corresponding experiment on coherent states
superposition with small mean photon numbers \cite{brune1}. In this
experiment the decoherence time comes out shorter than the photon
decay time of the cavity thus it may be taken as a good indication
of the mesoscopic nature of the superposition. We also wish to
observe that our state check procedure gives an unambiguous signal
when the state superposition is perfect. However if, because of
decoherence or state preparation, the final superposition is not
perfect, our procedure is yet able to measure the degree of
coherence (or the state preparation fidelity) of the 2QSB. Although
a detailed ab initio analysis is required for the general case, we
give here a quantitative simple example of this aspect. In fact, if
the initial state of the system leads to a final state of the form
given in the first line of Eq.~(\ref{probe3}) plus the term
$\delta\ket{\uparrow_1\downarrow_2}$ and with a new global
normalization factor $\mathcal{N}=1/\sqrt{2+|\delta|^2}$, the
probability of detecting the first atom in $\ket{\uparrow_1}$ and
the second atom in $\ket{\downarrow_2}$ is
$\mathcal{P}(\uparrow_1\downarrow_2)=(\mathcal{N}|\delta|)^2$ . So,
the state preparation fidelity is given by
$F=(1+|\delta|^2/2)^{-1}=1-\mathcal{P}(\uparrow_1\downarrow_2)$ and
it is therefore determined by the detection outcomes.

In this paper, we have defined the $N$-photon quantum superposition
of two orthogonal generalized binomial states of electromagnetic
field ($N$QSB). Our main result is the proposal of non-conditional
schemes to generate and reveal such an ``optimized'' quantum
superposition in a single-mode high-$Q$ cavity in the case $N=2$. We
wish to emphasize that the orthogonality of the two generalized
binomial states forming this state plays a crucial role to reveal,
by resonant probe atoms, the quantum nature of the superposition.
The implementation of the proposed schemes has been also analyzed,
showing how the unavoidable errors characterizing the current
experiments do not seem to sensibly affect them. Because of the
orthogonality property of generalized binomial states with any $N$,
our generation and revealing procedure of their quantum
superposition may be in principle extended to the cases with $N$
larger than two. This would lead, for $N\gg1$, to a regime of
macroscopic quantum superpositions, the highest value of $N$ being
only limited by the experimental capabilities. The results of this
paper can provide the basis for both new knowledge about the
foundations of quantum theory (measurement process,
quantum-classical border) and applications in quantum information
processing, in analogy with the superpositions of coherent states
\cite{jeong}.

\end{document}